\documentclass[aps,prd,twocolumn,nofootinbib,showpacs,superscriptaddress]{revtex4-1}

\usepackage{amsmath,amsfonts,amssymb,bm}
\usepackage{graphicx}
\usepackage{subfigure}
\usepackage{color}

\usepackage[colorlinks=true, pdfstartview=FitV, linkcolor=purple, citecolor= purple, urlcolor=blue]{hyperref}
\usepackage[nameinlink]{cleveref}

\definecolor{purple}{rgb}{0.5,0,0.5}
\definecolor{blue}{rgb}{0.0,0,0.9}

\newcommand{\mising}{M_{\text{Ising}}}
\newcommand{\partialD}[2]{\dfrac{\partial #1}{\partial #2}}

\graphicspath{ {./figure/} }

\newcommand{\tr}{{\text{tr}}}
\newcommand{\imag}{\text{i}}

\newcommand{\vtr}[1]{\boldsymbol{#1}}

\newcommand{\intelm}[2]{\mathrm{d}^{#1}#2} 

\usepackage{xcolor}

\begin{document}

\title{The equation of state and surface tension of  QCD in the first order phase transition region}

\author{Yuan-jun Mei}
\affiliation{School of Physics, Beijing Institute of Technology, Beijing 100081, China}

\author{Yi Lu}
\email{qwertylou@pku.edu.cn}
\affiliation{Department of Physics and State Key Laboratory of Nuclear Physics and Technology, Peking University, Beijing 100871, China}

\author{Fei Gao}
\email[]{fei.gao@bit.edu.cn}
\affiliation{School of Physics, Beijing Institute of Technology, Beijing 100081, China}

\author{Yu-xin Liu }
\email{yxliu@pku.edu.cn}
\affiliation{Department of Physics and State Key Laboratory of
Nuclear Physics and Technology, Peking University, Beijing 100871, China}
\affiliation{Center for High Energy Physics, Peking University, Beijing 100871, China}
\affiliation{Collaborative Innovation Center of Quantum Matter, Beijing 100871, China.}

\date{\today}

\begin{abstract}
We build up a complete description of QCD phase structure  by applying the parametrization of the chiral and deconfinement order parameters upon the calculations from functional QCD approaches.  In particular in the first order phase transition region at high chemical potential, both the phase transition line using Maxwell construction and the coexistence  boundary lines from the spinodal decomposition are determined.  We compute the thermodynamic quantities including the number density, the energy density, the pressure and also the free energy for  both stable  and   unstable phases of QCD. Additionally,  after  applying a phenomenological description of the inhomogeneity of the QCD free energy, we   obtain the surface tension of the first order phase transition of QCD. 
\end{abstract}


\maketitle

\section{Introduction}

The phase structure and thermodynamics of strong interaction matter at finite temperature $T$ and baryon chemical potential $\mu_B$ is  of great importance for understanding Quantum Chromodynamics (QCD). From the theoretical side,  it has been extensively studied via effective models and extrapolation methods~\cite{Shao:2011fk,Xin:2014ela,He:2013qq,Chelabi:2015gpc,Kojo:2020ztt,Chen:2020ath,Hippert:2023bel,Cai:2022omk,Chen:2018vty,Basar:2023nkp,Adam:2025pii,Ecker:2025vnb}, and first principle QCD approaches like  lattice QCD simulation~\cite{Borsanyi:2020fev,HotQCD:2018pds,Bonati:2018nut}, functional QCD (fQCD) methods including the Dyson-Schwinger equations (DSEs)~\cite{Qin:2010nq,Fischer:2014ata,Gao:2016qkh,Fischer:2018sdj,Gao:2020qsj,Gao:2020fbl,Gunkel:2021oya} and functional Renormalization Group (fRG) approach~\cite{Fu:2019hdw,Dupuis:2020fhh,Fu:2022gou}.  One of the main goals of these theoretical studies is to explain the experimental data from the Beam Energy Scan (BES) at the Relativistic Heavy-Ion Collider (RHIC)~\cite{Akiba:2015jwa,Luo:2017faz,HADES:2019auv,Lovato:2022vgq,ALICE-USA:2022glt,Arslandok:2023utm,Chen:2024aom}
. This will also need the assistance of non equilibrium physics like hydrodynamics simulations and dynamical evolution equations, where  the equation of state (EoS) of QCD serves as an essential input ~\cite{Ding:2017giu,Guenther:2020jwe,Monnai:2021kgu,Philipsen:2021qji,Karsch:2022opd,Ratti:2022qgf,Sorensen:2023zkk}.  Moreover,  The description of QCD phase transition, {as well as the  phase transition in other QCD-like theories,  is also important in cosmological studies as the cosmic  first order phase transition can be related to the gravitational wave (GW)  production through the thermodynamic quantities, particularly the trace anomaly and the surface tension of the system are required~\cite{Gao:2021nwz,Gao:2023djs,Gao:2024fhm,Zheng:2024tib,Reichert:2021cvs,Pasechnik:2023hwv,Cline:2025bwe}. The quantitative description is required for further constraining the potential role of such  phase transitions in the GW production.}
 
A direct precision calculation of the QCD thermodynamic quantities  has been extended to the chemical potential  region up to $\mu_B/T \lesssim 4$ ~\cite{Lu:2025cls}. This region remains the crossover region as the same calculation also suggests the critical end point is located at $\mu_B/T \sim 5-6$.  However,  for thermodynamic quantities in the first order phase transition region, there is still lack of precision calculations from first principle QCD approaches. Alternatively, one may utilize the Ising parametrization  to study the QCD thermodynamics by mapping  QCD system to Ising model~\cite{Parotto:2018pwx,Dore:2022qyz,Lu:2023msn,Karthein:2024zvs}.  One may apply the Ising mapping to map the order parameters of QCD instead of the thermodynamic quantities, and after that, one can incorporate the parameterized order parameters into the quark propagator to calculate the QCD thermodynamic quantities~\cite{Lu:2023msn}.  This method has  provided a quantitatively accurate  description of the QCD thermodynamics for low chemical potential, and has also enabled the description of  Maxwell construction in the first order phase transition region. Here in this article, we extend this method to provide a complete description of the first order phase transition,  incorporating the application of unstable phase description  from Ising model as introduced in Ref.~\cite{Karthein:2024zvs}.  The obtained trace anomaly and surface tension of QCD can open access for the precision phenomenological studies related to QCD first order phase transition both in hydrodynamics simulations and  the cosmological studies.

The article is organized as follows: 
In Sec.\,\ref{sec:med}, we present the framework of the parametrization of  order parameters and the construction of thermodynamic quantities.
In Sec.\,\ref{sec:num}, we show the numerical results of the constructed QCD phase diagram including the phase transition line and the coexistence boundary line, and then the respective QCD thermodynamic quantities,  in particular, the surface tension in the first order phase transition region.
In Sec.\,\ref{sec:sum}, we summarize the main results and make a few discussions.

\section{The general method}\label{sec:med}
In functional QCD approaches,   the quark propagator  is the central element in determining the thermodynamic quantities. The chiral and deconfinement order parameters are directly linked to the properties  of the propagator. As a result, parameterizing the order parameters and integrating   them into the quark propagator  offers a clear and detailed microscopic description of the system.   Additionally, this parameterization preserves higher-order information, even when a mean-field Ansatz is applied, as it effectively encapsulates the non-perturbative properties of QCD's correlation functions.
\subsection{From order parameters to QCD thermodynamics}
Generally speaking, the QCD phase transitions are characterized by  two order parameters, the chiral and the deconfinement phase transitions correspond to the chiral condensate $\langle \bar{q}q \rangle$ and the Polyakov loop $\Phi$, respectively.
Microscopically, the two order parameters are embedded in the Green functions, i.e. the full quark propagator (inverse) in the language of functional approaches:
\begin{equation}\label{eq:fullquarkprop}
\begin{split}
  S_{q}^{-1}(p) = & \, \imag (\tilde{\omega}_{n} +  gA_{4} )\gamma_{4} \, Z_q^E(\boldsymbol{p},\tilde{\omega}_n) \\
  & + \imag \boldsymbol{\gamma}\cdot\boldsymbol{p} \,  Z_q^M(\boldsymbol{p},\tilde{\omega}_n) + Z_q^E(\boldsymbol{p},\tilde{\omega}_n) \, M_{q}(\boldsymbol{p},\tilde{\omega}_n),
\end{split}
\end{equation}
where $\tilde{\omega}_{n} = \omega_n + \imag \mu_q$ with $\omega_{n}$ the Matsubara frequencies, $\mu_q$ the quark chemical potential and $\boldsymbol{p}$ the spacial momentum, along with $Z_q^E$, $Z_q^M$ and $M_q$ the dressing functions and $A_{4}$ the gluonic background field condensate which is related to the Polyakov loop $L$ as~\cite{Fister:2013bh,Fischer:2013eca}:
\begin{equation}\label{eq:LA0}
L[A_{4}] = \frac{1}{N_c} \tr \mathcal{P} e^{\imag \int_0^{\beta} d x_4 A_4}.
\end{equation}
with $\beta = 1/T$. The quantities in the quark propagator, $M_q$ and $A_4$, can be considered as  the order parameters of chiral and deconfinement phase transition, respectively.
 
We then applied  a further approximation on the dressing functions at each $(T,\mu_q)$ in Eq.~(\ref{eq:fullquarkprop}):
\begin{equation}\label{eq:propapprox}
\begin{split}
  & Z_q^{E,M}(\boldsymbol{p},\tilde{\omega}_n; T,\mu_q) \to 1, \\
  & M_q(\boldsymbol{p},\tilde{\omega}_n;T,\mu_q) \to M_q(0;T,\mu_q).
\end{split}
\end{equation}
This then leads to a simple expression of quark propagator as:
\begin{equation}\label{eq:fullquarkprop0}
\begin{split}
   S_{q}^{-1}(p) = \imag[\tilde{\omega}_{n} +  gA_{4} (T,\mu_q)]\gamma_{4} + \imag \boldsymbol{\gamma}\cdot\boldsymbol{p} \,  +  M_{q}(T,\mu_q),
\end{split}
\end{equation}
in the expression we have left out the momentum index in  $M_q$ and labeled the $T$ and $\mu_q$ dependence in $A_4$ and $M_q$ explicitly.  Note that this approximation  is just to take the quark mass in the infrared limit, which generally contains the main information of QCD in the nonperturbative regime. {We emphasize that  for a  direct calculation from functional  QCD methods,  one will need to acquire the correct  momentum dependence in the quark mass function and wave function, and only after this, one can describe the QCD phase transition.  However, here we do not aim at solving the  quark mass and wave function,  instead,  we only need the correct $T$ and $\mu_B$ dependence of   the  QCD thermodynamic quantities.  Taking the infrared limit  will not change the  $T$ and $\mu_B$ dependence of the order parameters, and consequently, it can still give a good description of QCD thermodynamics. This has been previously verified  as the obtained thermodynamic quantities via this approximation are in good agreement with the results from lattice QCD simulations~\cite{Lu:2023msn}.}

Such an approximation allows one to give an analytic expression of number density $n_{q}$ in terms of the two order parameters of QCD as:
\begin{gather}
n_{q}(T,\mu_{q}) = - T \sum_n \int \frac{\textrm{d}^{3}\boldsymbol{p}}{(2\pi)^{3}} \tr_{C,D}[ \gamma_4 S_{q}(\boldsymbol{p},\omega_{n})]\, \notag\\
= 2 N_{c} \int \frac{\mathrm{d}^{3}\boldsymbol{k}}{(2\pi)^{3}} \left[ f_{q}^{+}(\boldsymbol{k};T,\mu_{q}) - f_{q}^{-}(\boldsymbol{k};T,\mu_{q}) \right], \label{eq:nq_constit} 
\end{gather}
with
\begin{gather}
f_{q}^{\pm} = \frac{L(T,\mu_{q}) x_{\pm}^{2} + 2L(T,\mu_{q}) x_{\pm} + 1}{x_{\pm}^{3} + 3L(T,\mu_{q}) x_{\pm}^{2} + 3L(T,\mu_{q}) x_{\pm} + 1}, \\[1mm]
x_{\pm}(\boldsymbol{k};T,\mu_{q}) = \exp \left[ (E_q(\boldsymbol{k};T,\mu_{q}) \mp \mu_{q})/T \right], \notag\\[1mm]
E_{q}(\boldsymbol{k};T,\mu_{q}) = \sqrt{\boldsymbol{k}^{2} + M_{q}^{2}(T,\mu_{q})} .\notag
\end{gather}
With the knowledge of the two order parameters $M_q$ and $L$ at each temperature and chemical potential, the number density can be obtained. The pressure is then calculated  together with  $P(T,0)$ from lattice QCD simulation via the relation~\cite{Chen:2012zx,Isserstedt:2020qll,Gao:2021nwz}:
\begin{align}
	P(T,\mu_B) = P(T,0) +\sum_{q}  \int\limits_0^{\mu_q} \mathrm{d}\mu_q\, n^{\ }_q(T,\mu) \,.
	\label{eq:EoSchem}
\end{align}
 The other thermodynamic quantities can be further calculated. 
\subsection{The parameterization of  the chiral and deconfinement order parameters}

As mentioned above, a direct computation with precision  for  $M_q$ and $L$ has been conducted within the functional QCD method in   the chemical potential  region up to $\mu_B/T \lesssim 4$ ~\cite{Lu:2025cls}. Together with the results from the lattice QCD simulation, there exists some instructions for the parameterization of $M_q$ and $L$. 
First of all, one expects the QCD chiral phase transition line for a wide range of chemical potential  can be parametrized as:
\begin{equation}
  \frac{T_{c}(\mu_B=3\mu_q)}{T_{c}(0)} =  1 - \kappa \left( \frac{3\mu_q}{T_{c}(0)} \right)^{2} -\kappa_2  \left( \frac{3\mu_q}{T_{c}(0)} \right)^{4} + \cdots \, .  \label{eq:TcmuB} 
\end{equation}
For the (2+1)-flavor case, the consistent results (central averages) are $T_{c}(0) = 155\;$MeV, $\kappa = 0.016$, and $\kappa_2 =3.5\times10^{-4}$~\cite{Borsanyi:2020fev,HotQCD:2018pds,Bonati:2018nut,Fu:2019hdw,Gao:2020fbl}. 
{Note that the validity of  the parametrization   beyond  $\mu_{B} \approx  700\,\textrm{MeV}$ is lack of   evidence from direct calculations~\cite{Fu:2023lcm,Gao:2020fbl,Gunkel:2021oya,Fu:2019hdw}. Nevertheless,  there are several hints that the phase transition line should follow this parametrization, for instance, at zero temperature, the liquid gas transition at $\mu_B \approx 930$ MeV gives a constraint on the end point of the line~\cite{Gao:2025kzk}. The study of the thermodynamic quantities also suggests that the first order phase transition line should follow a quadratic behavior of $\mu_B$~\cite{Gao:2015kea}.}  Here we will then extend this parametrization to the whole temperature region, and adjust the parameters to yield the phase transition at zero temperature at $\mu_B=930$ MeV.  

Moreover, the functional QCD approaches also give the estimation of the location of the CEP at about $\mu_{B}^{\textrm{CEP}} \approx 600$ to $650\,\textrm{MeV}$~\cite{Fu:2023lcm,Gao:2020fbl,Gunkel:2021oya,Fu:2019hdw,Lu:2025cls}. Here we apply the CEP at $$(T^{\textrm{CEP}},\mu_{B}^{\textrm{CEP}})=(106,600)~\rm{MeV}.$$
This gives a general constraint on the order parameters of QCD  phase transitions. 

Now for the order parameter of chiral phase transition,  $M_q$, it is found that the functional QCD data~\cite{Fu:2019hdw,Gao:2020fbl} fits well with the feature of the 3D Ising parametrization~\cite{Parotto:2018pwx,Dore:2022qyz}.  Note that in principle, any parametrization works equivalently if it matches with the data from direct computation, here for simplicity, we choose such a 3D Ising parametrization. 
In the Ising parametrization, one can  parametrize  $M_q$ as:
\begin{equation}\label{eq:MT}
  M_{q}(T,\mu_{q}) = \frac{M_{0}}{2} \left[ 1 - M_{\mathrm{Ising}}(T,\mu_q) \right].
\end{equation}
 $M_{0} = 350\;$MeV is chosen as the typical mass scale for the light-flavor quarks suggested by the lattice~\cite{Bowman:2005vx,Oliveira:2016muq} and functional QCD~\cite{Fu:2019hdw,Gao:2021wun}.  $\mising$ is parametrized with the order parameter  potential as:
 \begin{equation}
    F(\mising, r, h) = -h\mising + \dfrac{1}{2}r\mising^2 + \dfrac{\lambda}{6}\mising^6\,,
\end{equation}
where $r = (T - T_c)/T_c$ is the reduced temperature and $h$ is the external magnetic field.  Note that the potential is different from the Ising mean field potential, as the fQCD computation suggests that {for the effective potential of order parameter $\phi$ (here is simply the  quark mass $\phi=M_q$),} the  $\phi^6$ term is dominant near the QCD phase transition region~\cite{Braun:2023qak}. This $\phi^6$ dominance is natural in a $3d$ field theory, as the unit of the field in $3d$ is the square root of the energy scale that is determined by the kinetic term, and hence, $\phi^6$ becomes relevant following the instruction of Wilsonian‘s renormalization group.    Such a potential could describe both the second order and first order phase transition including the spinodal line that gives the coexistence boundary as proposed in Ref.~\cite{Karthein:2024zvs}, and as will be illustrated below, the parameter $\lambda$ is the control parameter for the range of the coexistence region. 

The order parameter $\mising(r,h)$ can be obtained from the potential via:
\begin{equation}\label{eq:Landau_equilibrium}
    \partialD{F}{\mising} = -h + r\mising +\lambda \mising^5 = 0\,.
\end{equation}
In order to match the phase transition line Eq.~(\ref{eq:TcmuB}) precisely, one can take a non-linear mapping from $r,h$ to $T,\mu_B$ as:
\begin{equation}\label{eq:Ising_QCD_mapping}
    \begin{split}
        \dfrac{3\mu_q - \mu_B^\text{CEP}}{\mu_B^\text{CEP}} &= -r\omega\rho\cos\alpha_1 - h\omega\cos\alpha_2\,,\\
        \dfrac{T - T^\text{CEP}}{T^\text{CEP}} &= f_{PT}(r) + h\omega\sin\alpha_2\,,
    \end{split}
\end{equation}
where 
\begin{equation}\label{eq:Ising_QCD_mapping}
    \begin{split}
&\alpha_1 = \arctan\left[2\kappa\left(\dfrac{\mu_B^\text{CEP}}{T_c(0)}\right) + 
4\kappa_2 \left(\dfrac{\mu_B^\text{CEP}}{T_c(0)}\right)^3\right]\,,\\
& \alpha_2 = \alpha_1 + \dfrac{\pi}{2}\,,\quad
\omega = 1\,,\quad\rho = 2\,.
    \end{split}
\end{equation}
The function $f_{PT}(r)$ is chosen such that the mapping match the the chiral phase transition line . Hence, 
\begin{eqnarray}
   f_{PT}(r) =  &&\dfrac{T_c(0)}{T^\text{CEP}}{ \big [}1 - 
    \kappa\left(\dfrac{\mu_B^\text{CEP}}{T_c(0)}\right)^2(1 - r\omega\rho\cos\alpha_1)^2 \notag \\
&&   -\kappa_2 \left(\dfrac{\mu_B^\text{CEP}}{T_c(0)}\right)^4(1 - r\omega\rho\cos\alpha_1)^4 {\big ]} - 1\,.
\end{eqnarray}

For the Polyakov loop expectation value, we take the fRG result at zero chemical potential from Ref.~\cite{Fu:2015naa}, which is denoted as $\Phi(T,0) = L(t)$ with $t = T/T_c(0)$. The following fit function is taken here:
\begin{equation}
  L(t) = 2 \left[ 1 + \exp \left( \frac{1+l_1 t^3}{m_1 t + m_2 t^6} \right) \right]^{-1},
\end{equation}
with $l_1 = 2.732$, $m_1 = 0.5495$ and $m_2 = 1.831$.
For the Polyakov loop at finite chemical potential, we consider it along with the same phase transition line as the chiral phase transition and thus we have:
\begin{gather}\label{eq:PhiT}
  \Phi(T,\mu_q) = L(t_{\Phi}), \\
  t_{\Phi} = \frac{T}{T_{c}(0)}+\kappa \left(\frac{ \mu_{B}}{T_{c}(0)}\right)^{2}+\kappa_2 \left(\frac{\mu_{B}}{T_{c}(0)}\right)^{4} \, .\notag
\end{gather}

To sum up, the above parametrization of the $T$ and $\mu_B$  dependence of the $M_q$ and $L$  opens an access for analyzing the QCD phase structure and thermodynamic quantities, including the first order phase transition region. Note that here the parametrization of the Polyakov loop is relatively crude, however, based on the fQCD computation~\cite{Lu:2025cls}, the Polyakov loop contributes less to the lower order of the QCD thermodynamic quantities, such as pressure, number density and energy density. A self consistent calculation is required  to investigate higher order fluctuations and also other detailed information, which will be pursued in  future studies.

\begin{figure}[t]
    \centering
    \includegraphics[scale=0.4]{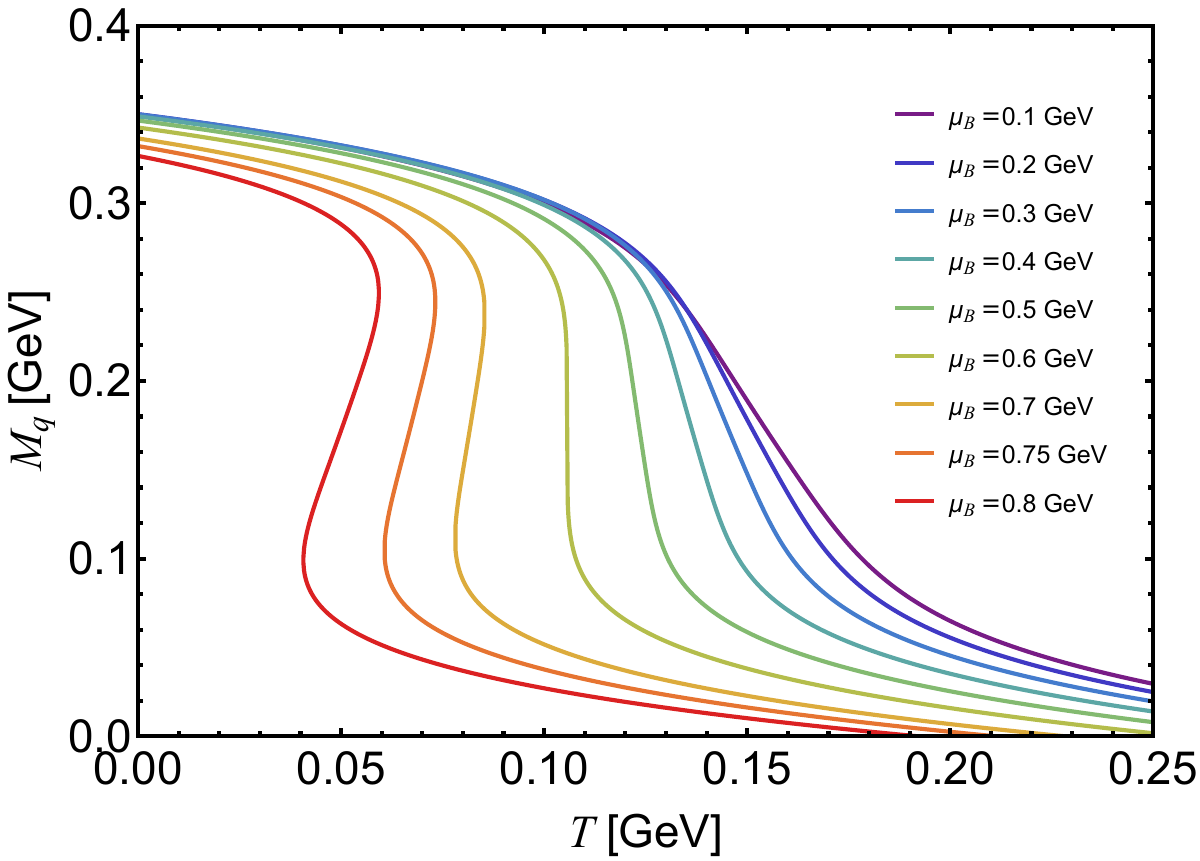}
    \caption{The temperature dependence of the order parameter of chiral phase transition at different chemical potential with the parameter in Ising potential being taken as $\lambda=1$. }
    \label{fig:Mq}
\end{figure}

\begin{figure}[t]
    \centering
    \includegraphics[scale=0.4]{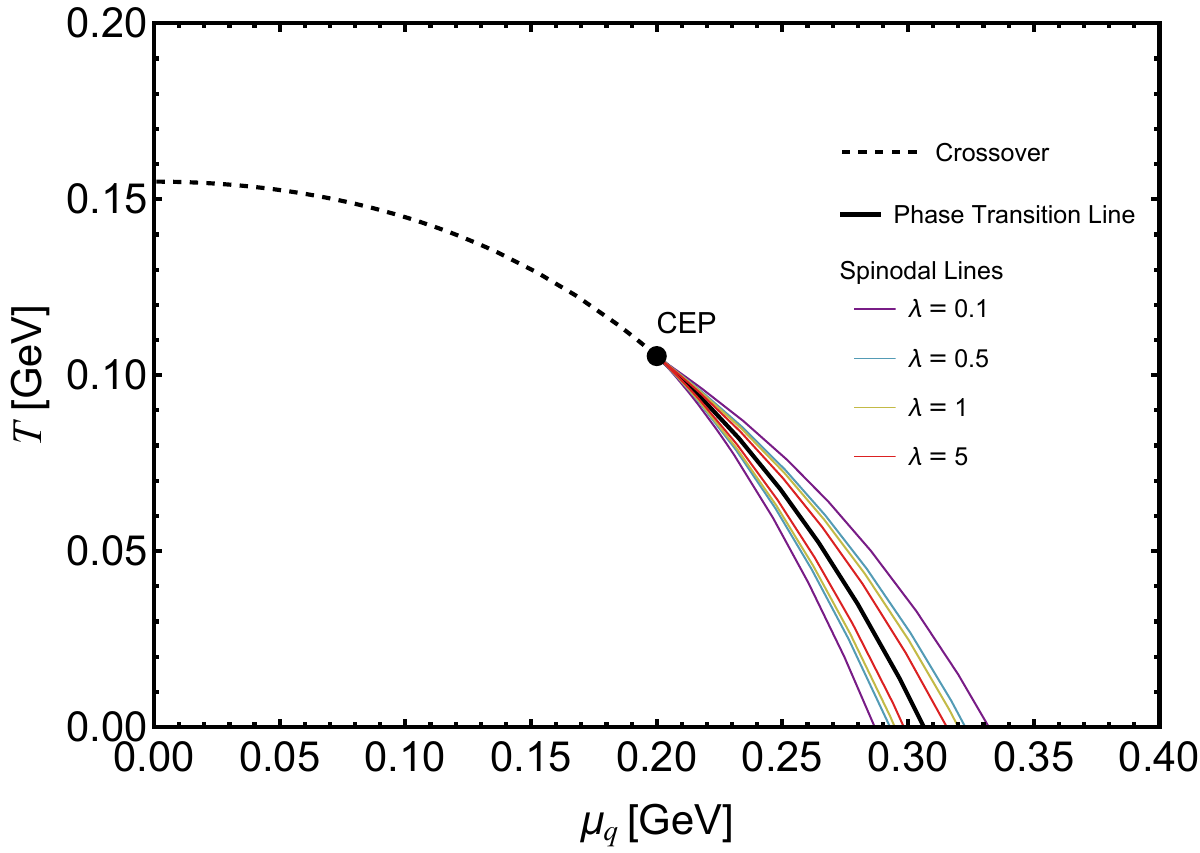}
    \caption{The phase diagram of chiral phase transition including the phase transition line and  the spinodal lines.}
    \label{fig:PT}
\end{figure}

\section{Numerical results}\label{sec:num}
In this section, we present the numerical results based on the above parametrization of $M_q(T,\mu_B)$ and $L(T,\mu_B)$. We  give the Chiral phase transition of QCD including the phase transition line and the coexistence boundary. We also compute  the thermodynamic quantities of QCD. Specifically, we calculate the pressure and the free energy in the first order phase transition, which then enables one to perform the  Maxwell construction, and analyze the supercooling/superheating phase. Furthermore, by incorporating  a phenomenological description of the inhomogeneity of QCD, we determine the surface tension during the QCD first order phase transition. 

\begin{figure}[t]
    \centering
    \includegraphics[scale=0.4]{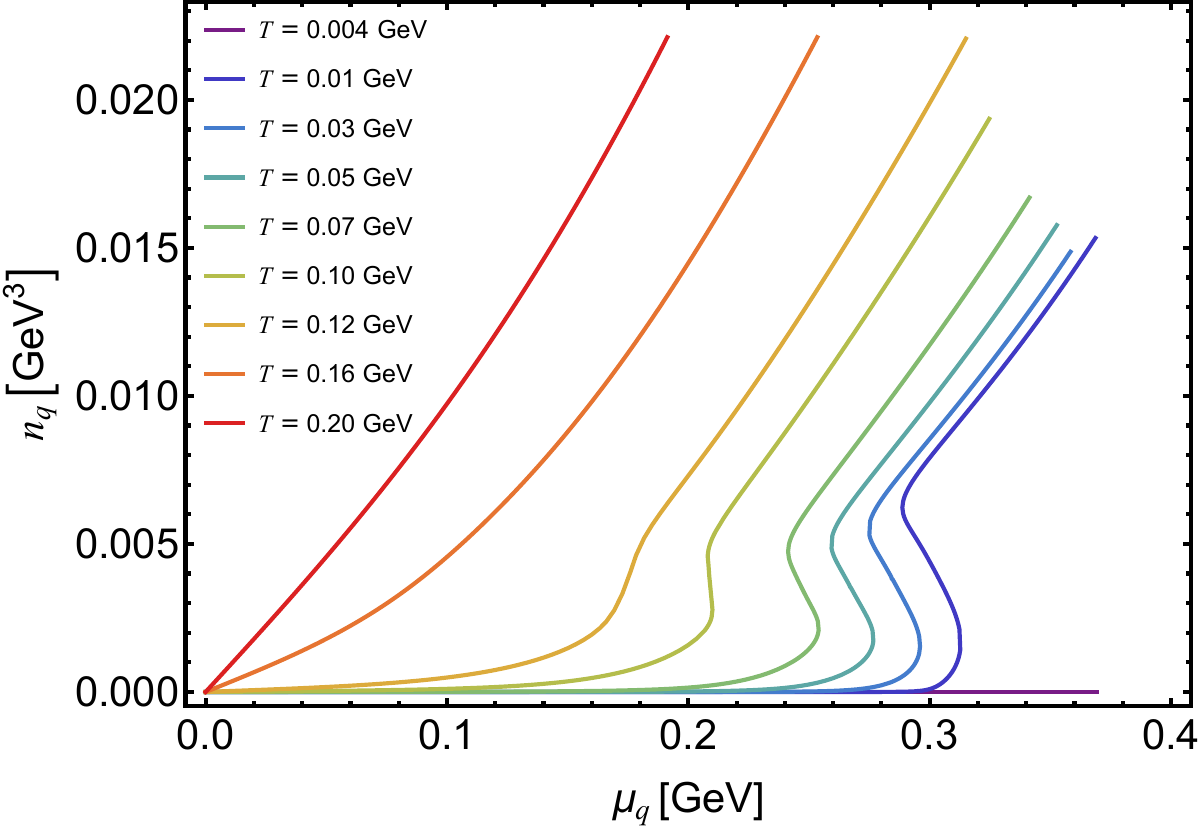}
    \caption{The number density as a function of $\mu_B$ at several temperatures, calculated from the quark propagator as in Eq.~\ref{eq:nq_constit}. }
    \label{fig:nq}
\end{figure}
\begin{figure}[t]
    \centering
    \includegraphics[scale=0.4]{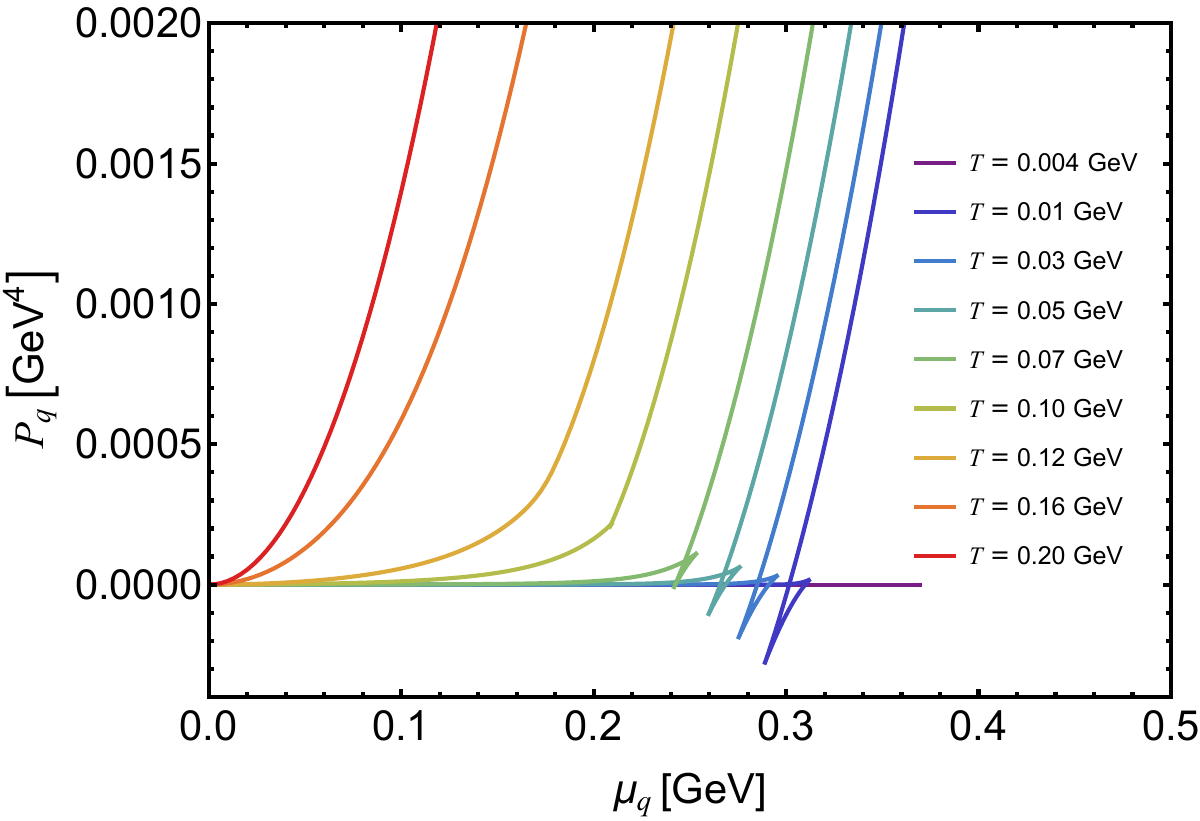}
    \caption{The pressure under the consideration of the metastable and unstable phase, calculated from Eq.~\ref{eq:Pnq}. }
    \label{fig:Pre}
\end{figure}

\begin{figure*}[t]
    \includegraphics[scale=0.418]{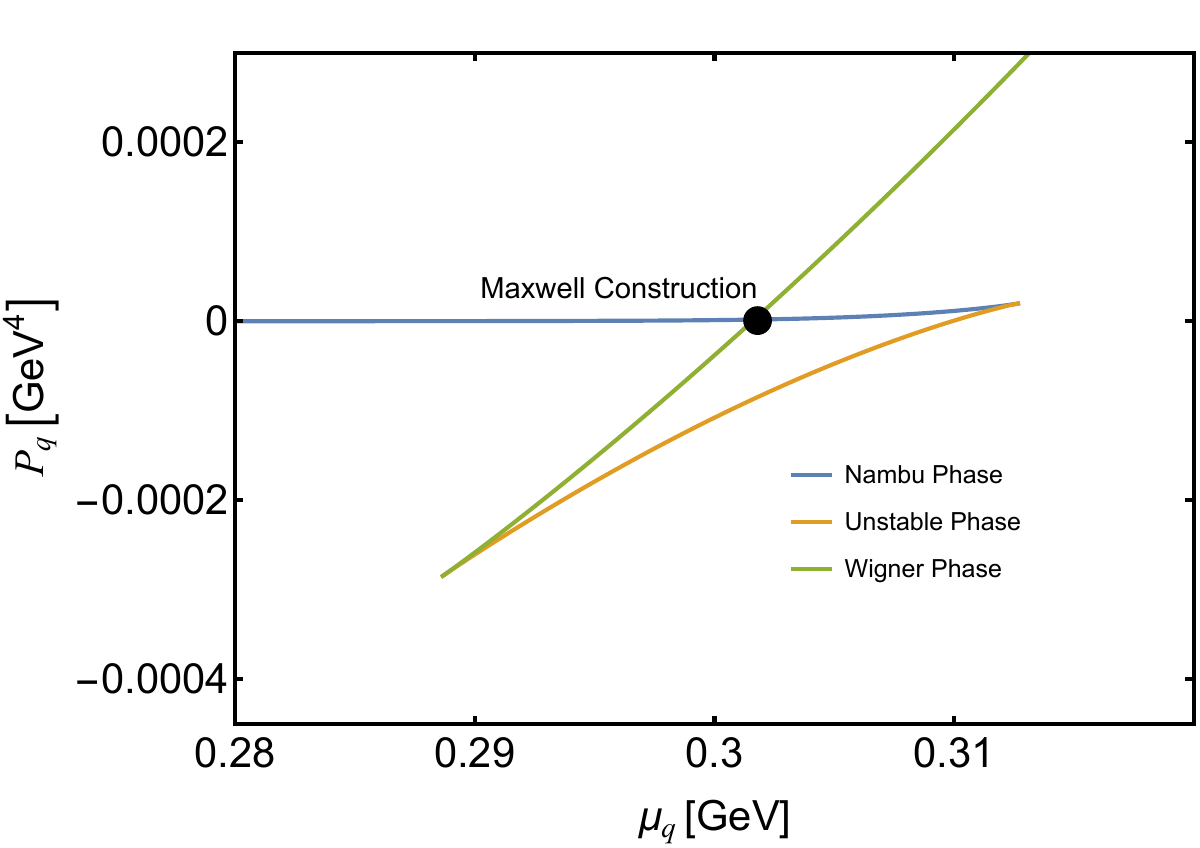} 
     \includegraphics[scale=0.4]{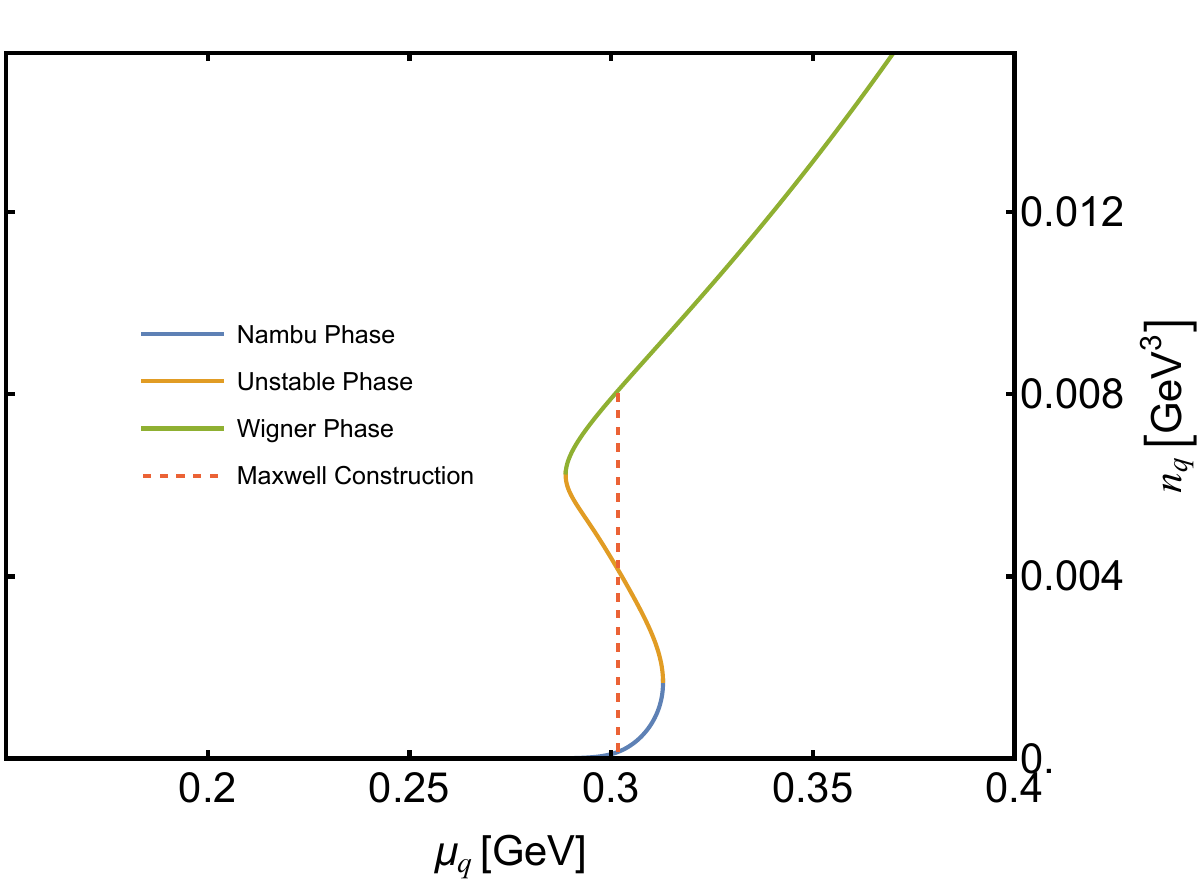}
    \caption{The Maxwell construction of the pressure at $T=0.01$ GeV (\emph{left panel}) and the respective number densities and the phase transition chemical potential (\emph{right panel}). }
    \label{fig:MC}
\end{figure*}

\subsection{QCD phase diagram}
Here we focus on the chiral phase transition, which dominants the thermodynamic properties of QCD for  temperatures below  approximately 300 MeV.  In the parameterization of the order parameter, the phase transition line has already been incorporated.  Therefore, the phase transition line can be directly read off from the chiral order parameter $M_q$, which its temperature and chemical potential dependence  is depicted in Fig.~\ref{fig:Mq}.  At low chemical potential, $M_q$ behaves as a monotonous  function, while as the chemical potential increases, the slope of the $M_q$ function becomes steeper. At the CEP location,  the slope becomes divergent. 

For larger chemical potential that the system steps into  the first order phase transition region, the function  $M_q$  becomes multi-valued, leading to  the bend over behavior in Fig.~\ref{fig:Mq}. For each temperature, the multi-valued function, in specific, has three values in the first order phase transition which corresponds to  three phases: the stable and the meta-stable phase, i.e., the chiral symmetry breaking phase (Nambu phase) and the chiral symmetry preserved phase (Wigner phase), and the unstable phase  located in between the stable and meta stable phase.   The upper and lower bounds of the chemical potential give the boundaries of the spinodal decomposition, thereby defining the coexistence region where the stable and metastable phases can coexist.  The respective phase diagram  is depicted in Fig.~\ref{fig:PT}.

As mentioned above, the range of coexistence region depends on the parameter $\lambda$. As the value of $\lambda$ decreases , the coexistence region becomes broader.  A further analysis for determining the parameter requires additional  information from the calculation of  functional QCD approaches in first order phase transition region, here we will adopt $\lambda=1$ to calculate the thermodynamic quantities. Note that the phase transition line in the first order phase transition is directly from the parameterization of Eq.~\ref{eq:TcmuB},  instead of being determined from  the thermodynamic quantities which requires the Maxwell construction from the pressure. This will be illustrated below that is in accordance with the direct parameterization.    {Such a  phase transition line  is the ideal case of phase transition, with the pressures of the stable phase and the meta stable phase being equal.  If considering the system evolves from high temperature to low temperature, the phase transition can happen  in  the coexistence region  at a lower temperature  after the evolution crosses the phase transition line, this is the so-called supercooling phenomenon.  Similarly for the opposite evolution direction,  the superheating phenomenon can take place. As the spinodal lines define the boundary of the coexistence region, they are also the lower and upper bound of the supercooling/heating phenomenon, respectively.}

\subsection{QCD Thermodynamic quantities}

After employing Eq.~\ref{eq:nq_constit} together with the parameterization of two order parameters in Eq.~\ref{eq:MT} and Eq.~\ref{eq:PhiT},  one can directly calculate  the quark number density in the temperature-chemical potential plane as depicted in Fig.~\ref{fig:nq}. For large temperatures that correspond to  the crossover region, the number density  increases monotonously as the chemical potential becomes larger. When the temperature is below the CEP temperature,   a first order phase transition occurs at a specific chemical potential. Similar to the behavior observed in $M_q$, here the complete parameterization that includes the metastable and unstable phases results in  a bend-over behavior in the number density.  This multi-valued region defines the coexistence region of the first order phase transition and the boundaries represent the spinodal lines. 

To determine the phase transition line, one can compute the pressure upon the integration of number density, and apply the Maxwell construction which yields the pressure of Nambu and Wigner phases to be equal at a certain chemical potential , i.e. $P_N(\mu_q)=P_W(\mu_q)$.  Note that since the number density is multi-valued in $\mu_q$, the integral respective to $\mu_q$ is ill defined, one can apply the intrinsic coordinate on the trajectory of the number density. The number density and $\mu_q$ can be then defined with the intrinsic coordinate as $n_q(T,\,s)$ and $\mu_q(s)$. The difference of the  pressure $P_q$ between finite $\mu_q$ and $\mu_q=0$, which is   the integral that follows the trajectory of function $n_q$ can be defined along with  $s$ as:
\begin{equation}\label{eq:Pnq}
    P_q(T,\boldsymbol{\mu}) =\sum_q \int_{0}^{{\mu}_{q}} n_{q}(T, s) \frac{\partial \mu_q(s)}{\partial s} \, \textrm{d} s.
\end{equation}
The pressure of the Nambu and Wigner phase, as well as  the unstable phase,  can be calculated in our parameterization. The pressure as a function of chemical potential at different temperatures has been depicted in Fig.~\ref{fig:Pre}.  In the first order phase transition,  the pressure  is also multi-valued. In general, the pressure of the stable phase is larger than that of  the meta stable phase, and  the Nambu and Wigner phase intersect at one  specific chemical potential which  precisely corresponds to  the phase transition point under the Maxwell construction.  

In Fig.~\ref{fig:MC}, we illustrate  the details of the Maxwell construction.   When the Nambu and Wigner phase  exceed the phase transition chemical potential, they become meta-stable phases, representing the  superheating/cooling phenomenon.  The line that connects the two phases is the pressure of unstable phase.  The instability of this phase  can be read off from its slope that corresponds to  a negative susceptibility. This is further  illustrated  clearly in the plot of number density  in the right panel of Fig.~\ref{fig:MC}. Besides, the dotted line is the   phase transition line under the Maxwell construction with the ending point value being the corresponding number density for the Nambu and Wigner phase.

\begin{figure}[t]
    \centering
    \includegraphics[scale=0.4]{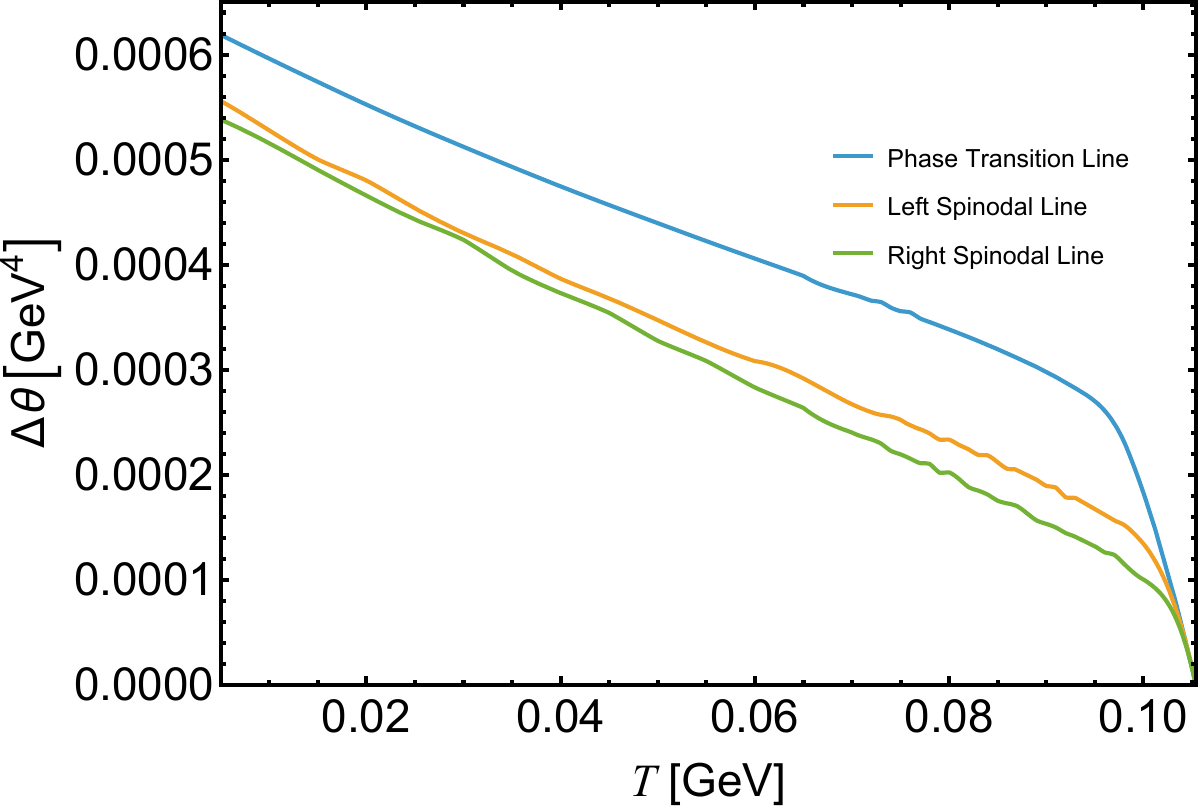}
    \caption{The  trace anomaly   { for one flavor quark}  along the phase transition line and also the spinodal lines. }
    \label{fig:Tr}
\end{figure}

{After having the pressure $P_q$, one can also calculate   the trace anomaly difference between two phases  in the coexistence region.  The trace anomaly difference denoting as $\Delta \theta$ can be  defined as:
\begin{eqnarray}\label{eq:traceanomaly}
 \Delta\theta&=&\Delta  P_q-\frac{1}{4}T\Delta  s_q-\frac{1}{4}\mu_q \Delta n_q\\
 &  =&-\frac{1}{4}T^5\frac{\partial   \Delta  P_q/T^4}{\partial T}-\frac{1}{4}\mu_q\frac{\partial  \Delta P_q}{\partial \mu_q},\notag
\end{eqnarray}
with $\Delta s_q$ the quark entropy density difference, $\Delta n_q$ the number density difference. The results of trace anomaly are depicted in Fig.~\ref{fig:Tr}. As the temperature decreases, the trace anomaly difference becomes larger which is due to the increasing chemical potential and the number density.}

\begin{figure}[t]
    \centering
    \includegraphics[scale=0.4]{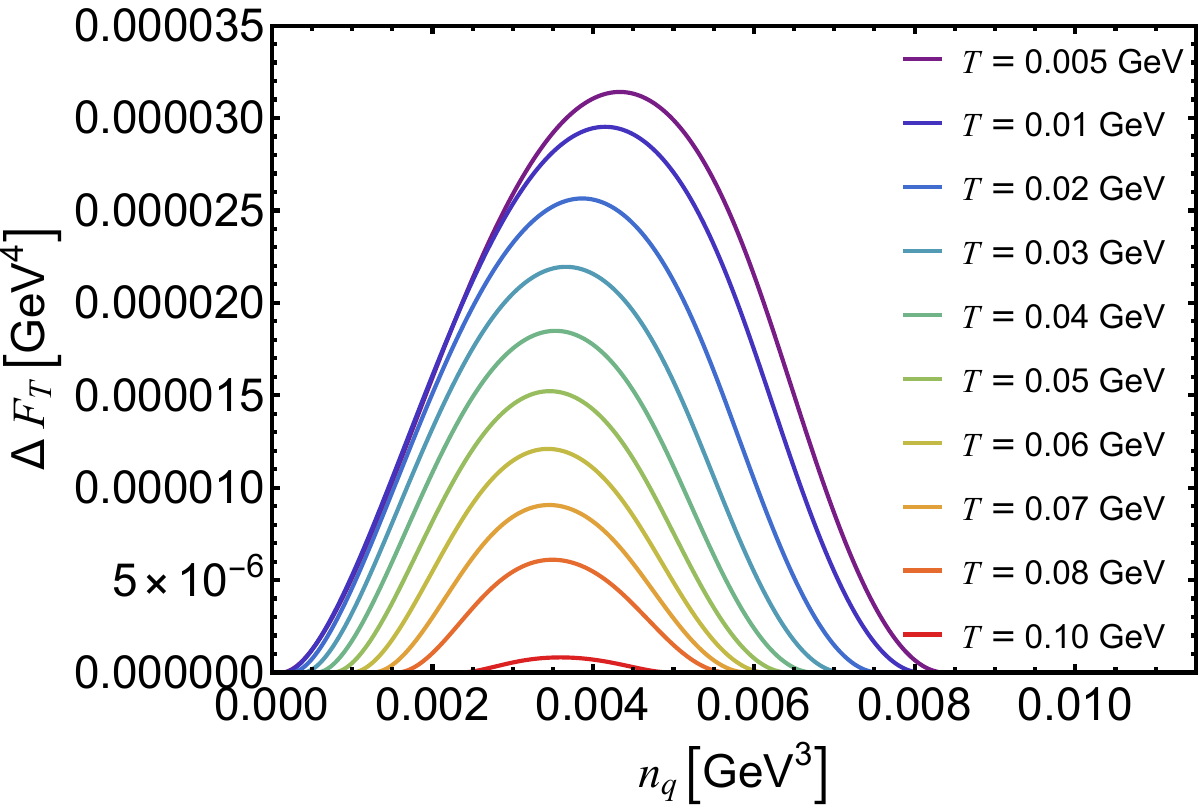}
    \caption{The deviation of free energy { for one flavor quark}  from the Maxwell construction at the phase transition points. }
    \label{fig:dFE}
\end{figure}

\begin{figure}[t]
    \centering
    \includegraphics[scale=0.4]{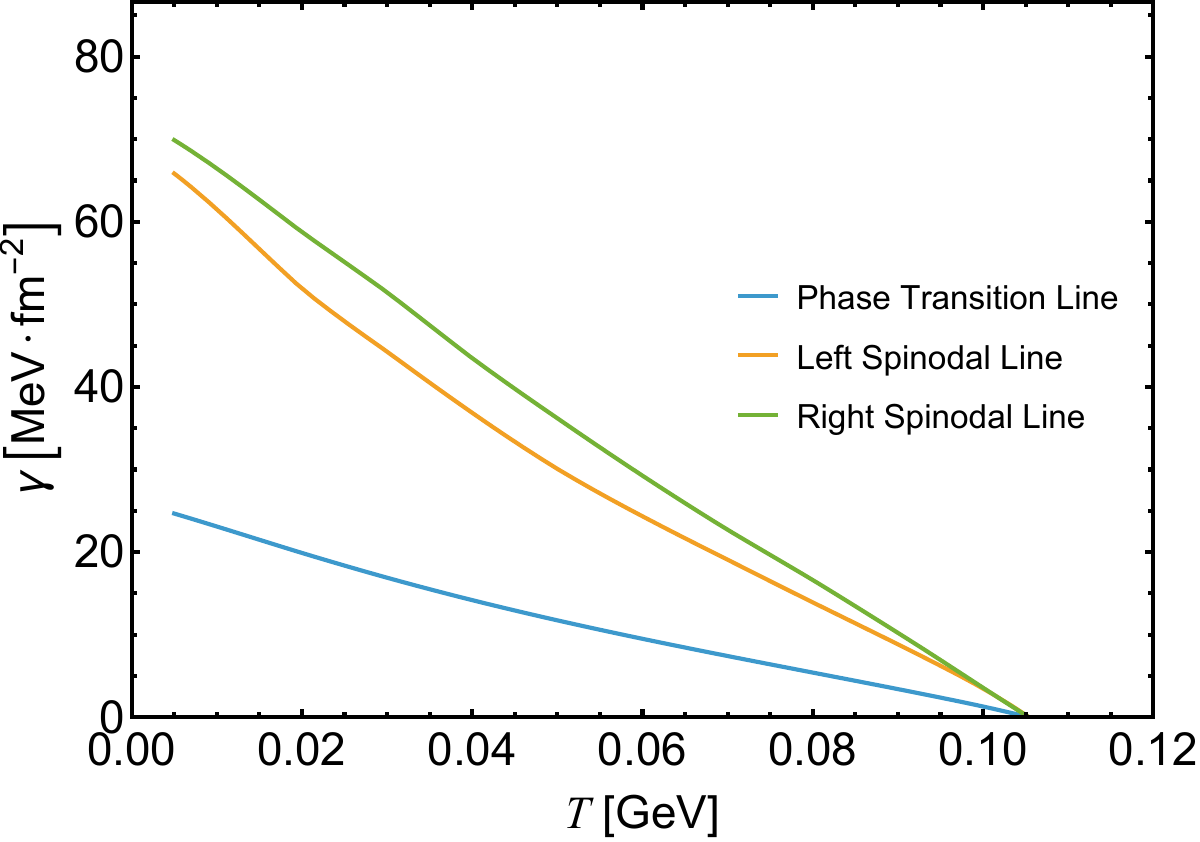}
    \caption{The surface tension  { for one flavor quark}  along the phase transition line and also the spinodal lines. }
    \label{fig:gamma}
\end{figure}

\subsection{Free energy and surface tension}

\begin{figure*}[t] 
    \includegraphics[scale=0.25]{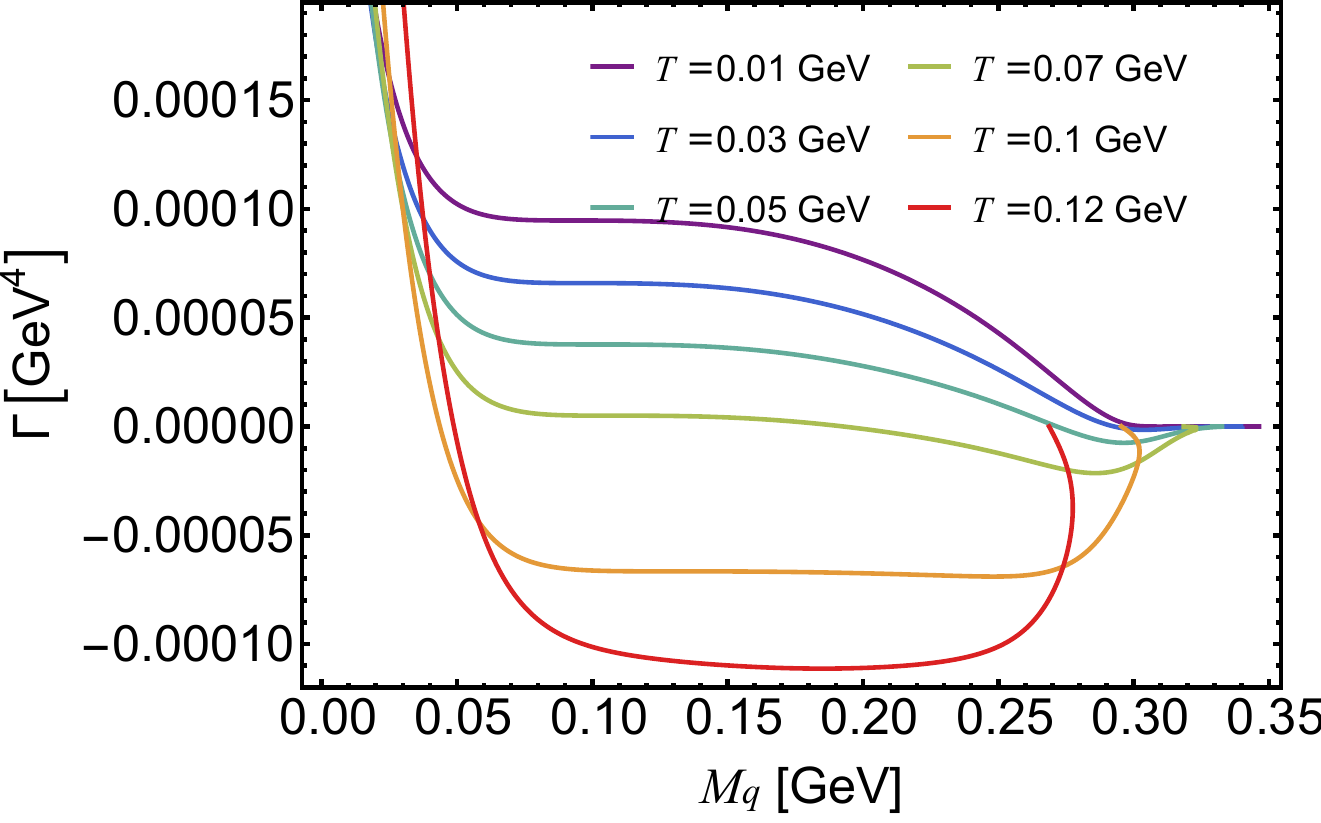}
        \includegraphics[scale=0.25]{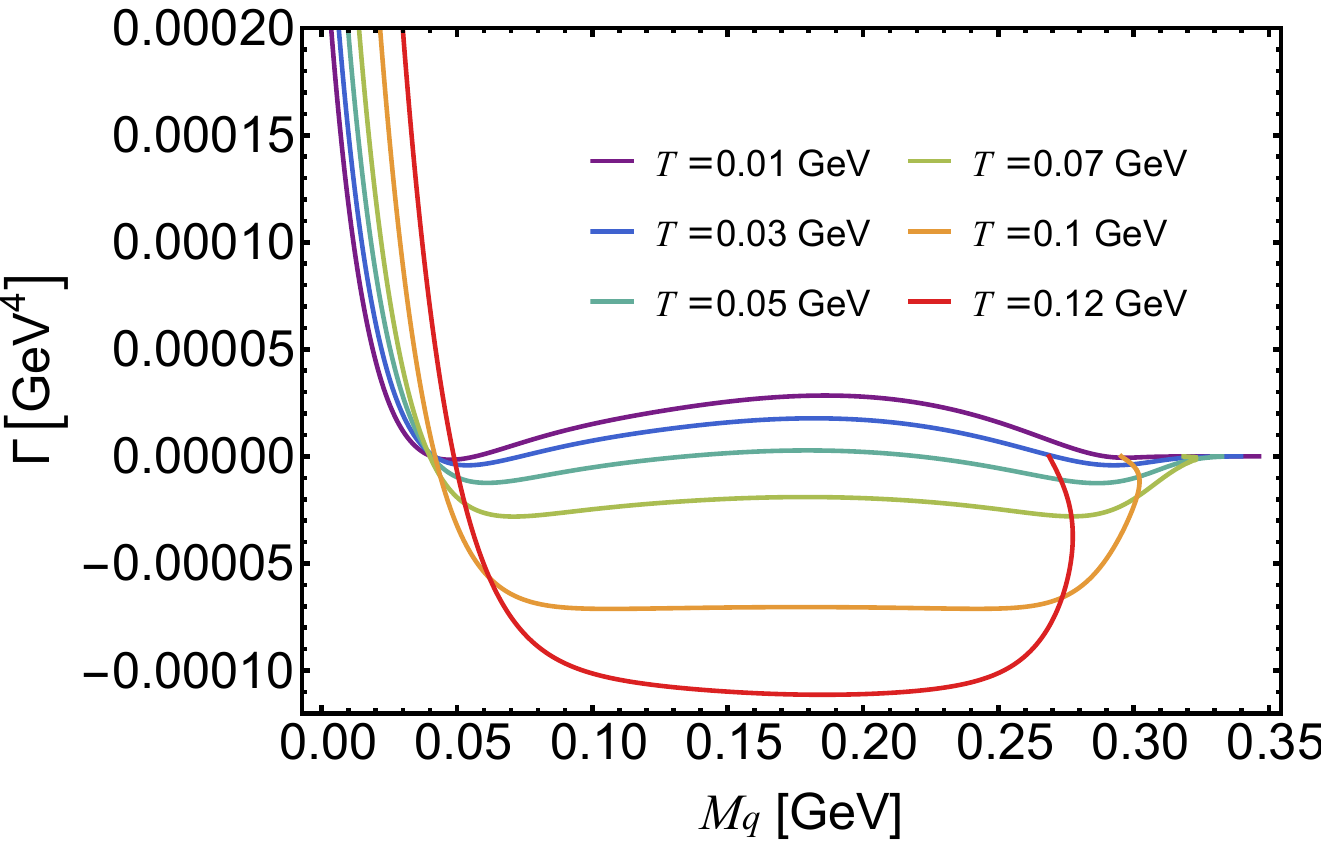}
            \includegraphics[scale=0.25]{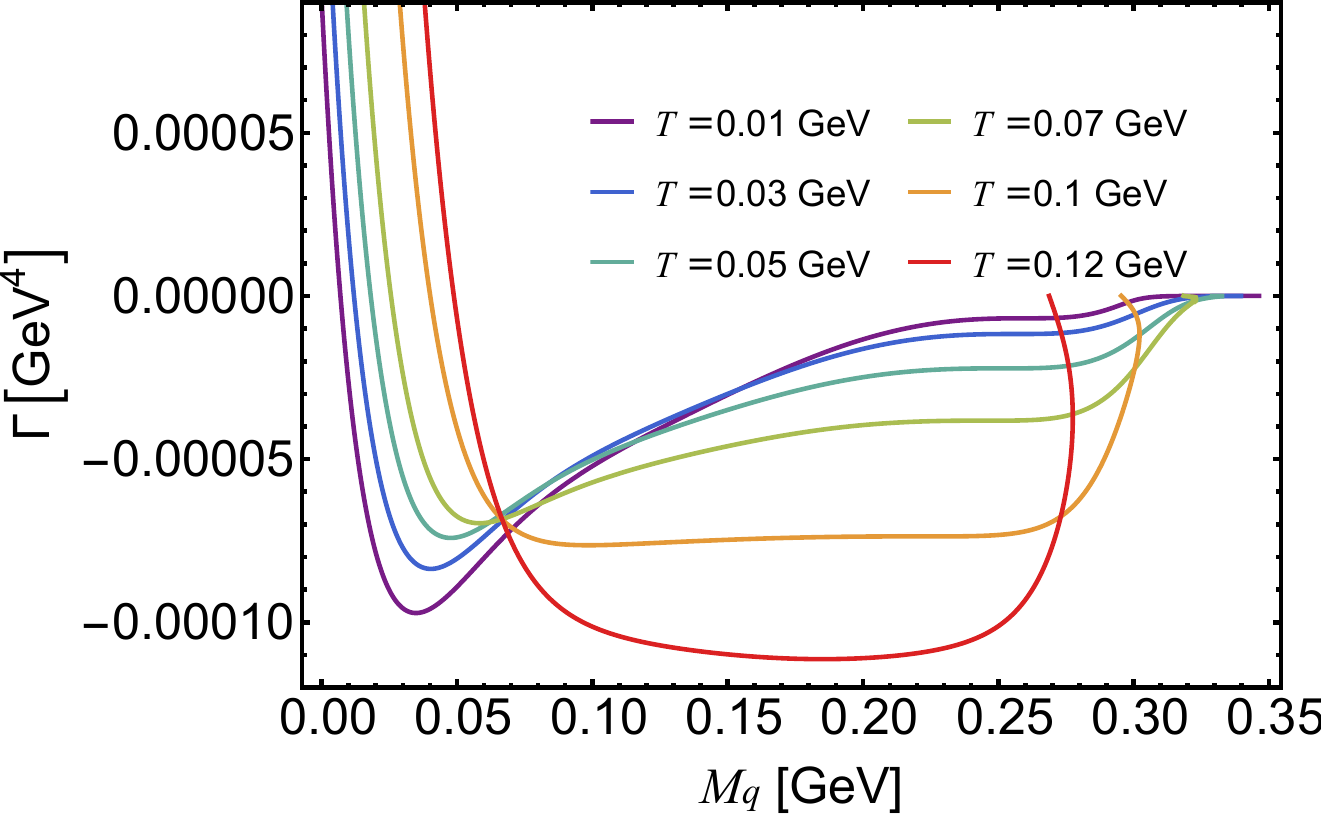}
    \caption{The effective potential  of Chiral phase transition  as the function of the order parameter $\phi=M_q$ at the phase transition point and also the spinodal boundaries.  At each temperature, the corresponding chemical potential in three figures increases from left to right.  }
    \label{fig:effpo}
\end{figure*}
One can also  calculate the free energy of the system using 
\begin{equation}\label{eq:Pnq}
    F_T(T, \,n_q) = F_T(T, 0) +   \int_{0}^{{n}_{q}} \mu_q(T,n_{q})\textrm{d}  n_{q} .
\end{equation}

It is more importantly to see the difference between the free energy of the system and the free energy under the Maxwell construction, which can be defined as:
 \begin{eqnarray}\label{eq:deltaFT}
    \Delta F_T(T, \,n_q) =  F_T(T, \,n_q) -   F_T^M(n_q),
\end{eqnarray}
where the  Maxwell construction $F_T^M(n_q)$ is the linear construction of free energy as:
 \begin{eqnarray}
 F_T^M(n_q) = F_T(n_N) + \dfrac{F_T(n_N) - F_T(n_W)}{n_N- n_W}(n_q - n_N).
  \end{eqnarray}
  The slope of the above Maxwell construction is precisely the chemical potential at phase transition.  The corresponding results for difference free energy at several temperatures are depicted  in Fig.~\ref{fig:dFE}.

The results of the difference free energy can be further incorporated to calculate the surface tension of the system. {A phenomenological inhomogeneous free energy
density for the first order phase transition system has been proposed~\cite{Randrup:2009gp},  which includes a gradient term to characterize the space distribution of the number density. The gradient term serves as the kinetic term of the system and determines the distribution of number density together with the free energy as the potential, and} we have
\begin{equation}
    F_T(\vtr{r}) = n_q\mu_q + \dfrac{1}{2}C(\nabla n_q)^2\,,
\end{equation}
where $C = \frac{a^2}{n_B^2}E_g$ with $n_B=\frac{1}{3}n^{\rm{CEP}}_q=0.19$ fm$^{-3}$ and $E_g=0.572$ GeV$\cdot \rm{fm}^{-3}$ to be the baryon number density and the energy density at the 
CEP. For the parameter $a$, the measure of the interface's thickness, we choose it to be $0.33\;$fm as that in Ref. \cite{Ke:2013wga,Gao:2016hks}.

By comparison with Eq.~\ref{eq:deltaFT}, one immediately gets:
\begin{equation}
    \Delta F_T -\dfrac{1}{2}C\left(\partialD{n_q}{r}\right)^2 = 0, 
\end{equation}
with the spherical symmetry being adopted.  The radial distribution of the number density and also the free energy can be calculated. 
The interface tension  is then expressed as:
\begin{equation}
    \gamma(T) = \int_{0}^{\infty}\Delta F_T(r)\,\intelm{}{r} = \int_{n_N}^{n_W}\sqrt{\dfrac{C}{2}\Delta F_T(n_q)}\,\intelm{}{n_q}\,.
\end{equation}
where $n_N$ and $n_W$ are number densities of the coexisting  Nambu and Wigner phase.   The corresponding results of surface tension are depicted in Fig.~\ref{fig:gamma}.  Note that such a calculation can be done in the whole coexistence region. In detail, the surface tension on the phase transition line gives a lower bound, while at the spinodal line, the surface tension is larger. A larger surface tension leads to a quicker phase transition which  is compatible with the supercooling/heating phenomenon. 

{Moreover, the effective potential of QCD as a function of the order parameter of chiral phase transition $\phi=M_q$ can be defined by utilizing the relation between $n_q$ and $M_q$. The effective potential $\Gamma(\phi)$ is  expressed as:
\begin{equation}
\Gamma(\phi=M_q;T,\mu_q)=F_T(T, n_q(M_q))-\mu_q n_q(T,M_q).
\end{equation}
The results of the effective potential are depicted in Fig.~\ref{fig:effpo}. Below the CEP temperature, the effective potential  gives the typical  "double-well" structure as a signal of the first order phase transition. At the phase transition point under the Maxwell construction,   the two minima  become equal. On the spinodal boundaries,  the minimum point for the meta stable phase becomes a saddle point. The effective potential of the order parameter can be further applied for studying the dynamical evolution of the nucleation and also  the spindonal decomposition process during the QCD phase transition.}

{ It is worth mentioning that the current calculations can be directly applied to calculate the GW spectral.  In particular,    the parameter $\alpha$ can be directly calculated from the  trace anomaly between two phases, as $\alpha\sim \theta$. The trace anomaly depends on the flavor, for example, a factor two should be involved as in the conventional QCD phase transition,  the up and down quarks experience the QCD phase transition simultaneously.  The parameter $\beta$ is related to the dynamics of phase transition which is in principle determined by the order parameter potential  in Fig.~\ref{fig:effpo}.  Such a  determination  is beyond the scope of this work,  however,  the parameter $\beta$ can be estimated via the surface tension as in Ref.~\cite{Cline:2025bwe} with  the relation between $\beta$ and the surface tension as $\beta\sim \gamma^{-3/2}$.       }

{We would also like to emphasize that  the GW signals obtained from the QCD phase transition in standard model, are very unlikely to be detected~\cite{Zheng:2024tib,Reichert:2021cvs,Cline:2025bwe}.  In specific, the surface tension of QCD is too small which then gives a large value of $\beta$ and cannot generate the strong enough GW signals. However, if incorporating  the scenario beyond standard model, either by a dark QCD sector, or in some scenarios, the quark mass and the coupling strength can be altered,  the situation will be different.   Our  frame here can be easily adopted to study  these QCD-like scenarios,  by adjusting the phase transition line and CEP location, as well as  the parameterization of order parameters, and hence, we offer a platform for calculating the GW signals from the cosmic first order phase transition of QCD-like theories. }
\section{summary}\label{sec:sum}
We develop a unified framework to describe the QCD thermodynamics upon the numerical results from lattice QCD simulations and the functional QCD approaches.  The key  is to retreat the quark mass and the gluonic background condensate as the order parameters of chiral and deconfinement order parameters respectively. On one hand,  these two elements  can be calculated directly through the functional QCD approaches and  easily parameterized. On the other hand, QCD thermodynamic quantities are microscopically encoded in these elements through the quark propagator, providing a bridge between first principle calculations and broader phenomenology studies.

Using this framework, we  calculate the QCD thermodynamic quantities encompassing stable, meta-stable phase and also the unstable phase.  This first completes the phase diagram of QCD in particular in the first order phase transition. The phase transition line can be determined through the Maxwell construction, and the coexistence region with the spinodal lines are also determined. 

A more interesting extension is the calculation of the  free energy of the system and its deviation from that of the Maxwell construction. This provides insights into the dynamics of  the first order phase transition. Moreover,  with the assist of the phenomenological description of the inhomogeneity, one can calculate the space distribution of the free energy and the number density. The surface tension is then determined along both the phase transition line and the spinodal lines, offering valuable insights into phenomena observed in heavy-ion collision experiments and cosmological studies related to QCD phase transitions.

\section{Acknowledgements}
 FG and YL thank the other members of the  fQCD collaboration~\cite{fQCD}  for discussions and collaboration on related subjects. FG is supported by the National  Science Foundation of China under Grants  No. 12305134.
 This work is also supported by the National Natural Science Foundation of China under Grants  No. 12247107 and No. 12175007.  
\bibliography{foptmap.bib}

\end{document}